# High-throughput injection-acceleration of electron bunches from a linear accelerator to a laser wakefield accelerator

Yipeng Wu,[1,2] Jianfei Hua,[1,*] Zheng Zhou,[1] Jie Zhang,[1] Shuang Liu,[1] Bo Peng,[1] Yu Fang,[1] Xiaonan Ning,[1] Zan Nie,[2] Fei Li,[2] Chaojie Zhang,[2] Chih-Hao Pai,[1] Yingchao Du,[1,†] Wei Lu,[1,‡] Warren B. Mori,[2] and Chan Joshi[2]

[1]Department of Engineering Physics, Tsinghua University, Beijing 100084, China

[2]University of California Los Angeles, Los Angeles, California 90095, USA

Plasma-based accelerators (PBAs) driven by either intense lasers (laser wakefield accelerators, LWFAs)[1] or particle beams (plasma wakefield accelerators, PWFAs)[2], can accelerate charged particles at extremely high gradients compared to conventional radio-frequency (RF) accelerators. In the past two decades, great strides have been made in this field[3-10], making PBA a candidate for next-generation light sources and colliders[11]. However, these challenging applications necessarily require beams with good stability, high quality, controllable polarization and excellent reproducibility[12,13]. To date, such beams are generated only by conventional RF accelerators. Therefore, it is important to demonstrate the injection and acceleration of beams first produced

* jfhua@tsinghua.edu.cn
† dych@tsinghua.edu.cn
‡ weilu@tsinghua.edu.cn

**using a conventional RF accelerator, by a PBA. In some recent studies on LWFA staging and external injection-acceleration in PWFA only a very small fraction (from below 0.1% to few percent) of the injected charge (the coupling efficiency) was accelerated[8,9]. For future colliders where beam energy will need to be boosted using multiple stages, the coupling efficiency per stage must approach 100%. Here we report the first demonstration of external injection from a photocathode-RF-gun-based conventional linear accelerator (LINAC) into a LWFA and subsequent acceleration without any significant loss of charge or degradation of quality, which is achieved by properly shaping and matching the beam into the plasma structure. This is an important step towards realizing a high-throughput, multi-stage, high-energy, hybrid conventional-plasma accelerator**.

Multi-stage plasma accelerators are inherently difficult to build because of the micrometer-size beams and wake structures, exceedingly large focusing fields and the temporal synchronization precision required at the femtosecond scale. This is why in spite of some success, a high overall coupling efficiency or charge throughput of the externally injected beam has been difficult to achieve. Recently, staged acceleration of LWFA has been demonstrated at BELLA of LBNL[8]. Although a plasma lens is installed between two LWFA stages to refocus the electron beam exiting the first stage into the second stage, the overall coupling efficiency is typically 3.5% due to imperfect matching of focusing. Acceleration of conventional RF accelerator-produced electron beam in a self-modulated proton beam driven PWFA has recently been demonstrated at CERN[9]. Due to the complex interaction and the non-collinear matching between the long (several plasma wavelength) electron bunch and the plasma wake, the coupling efficiency is below 0.1%. On the other hand, although the two-electron bunch PWFA experiment at FACET of SLAC has achieved roughly 25% coupling efficiency, the rest of the ~75% of the charge was lost due to the oversized width and length of the electron beam, and the non-ideal matching of the injected bunch to the PWFA[10]. In all the above scenarios, such low coupling efficiencies will be problematic for certain practical applications of PBAs, such as a multi-stage collider, where the coupling efficiency must be near 100% per stage, otherwise the beam charge throughput will be significantly affected.

In order to reach 100% coupling efficiency, a good matching of the injected beam to the plasma wake, not only of the beam size and the wake size, but also of the beam phase-space and the wake structure, is critical. Such good matching is also essential for beam emittance preservation. Previous theoretical studies have suggested using longitudinally tailored density profiles to fulfill this challenging requirement[14,15]. However, there has not been any experimental demonstration so far. Here we demonstrate successful 100% coupling of a high-quality externally injected beam from a LINAC into a PBA and find conditions under which there is little growth of the beam energy spread or divergence during the injection-acceleration process.

The schematic layout of the experiment is shown in Fig. 1. A 31.3±0.05-MeV electron bunch from a photocathode-RF-gun-driven LINAC[16,17] co-propagates with an ultrashort ($40 \pm 2$ fs FWHM), energetic ($600 \pm 14$ mJ), 0.8-μm-wavelength laser pulse, which is focused to a spot size of $12.2\pm0.3$μm (a radius where the laser intensity is $1/e^2$ compared to the on-axis value) that contains ~55% of the laser pulse energy (Fig. 1a, giving the laser Rayleigh length $Z_R \approx 584\mu m$ and peak intensity $I_0 \approx 4\times10^{18} W cm^{-2}$). The laser pulse creates a plasma in a 6-mm-long helium gas jet and excites a wake throughout the plasma.

In order to achieve high coupling efficiency the plasma wake must be wider than the electron beam. The beam is tightly focused to a spot as small as $20.3\pm0.9$μm RMS (Fig. 1b) at the entrance of the plasma through optimization of the beam normalized emittance (~1mm mrad) and the beam focusing magnets. The drive laser focal position $z_f$ is scanned from 3.5mm ($6.0Z_R$) to 5.5mm ($9.4Z_R$) before the entrance of the plasma such that the laser expands in vacuum to a transverse size from $74.2\mu m$ to $115.3\mu m$ when it enters the plasma (the transverse wake size $\simeq$ the transverse laser size). In order for the wake wavelength to be longer than the injected beam, the beam is strongly velocity compressed to ~13fs RMS (near flat-top current profile) in the photocathode-RF-gun by properly launching the beam at low phase and reducing the bunch charge (~20fC) [18]. The plasma density (plateau value), $n_p$, is adjusted between $(2-6)\times10^{17} cm^{-3}$, such that a relatively long plasma wavelength $\lambda_p \approx 3.3\times \frac{10^{10}}{\sqrt{n_p[cm^{-3}]}}\mu m \approx 74.6\mu m - 43.1\mu m$ (248.7fs-143.7fs) can be obtained. This is also beneficial for reducing beam energy spread growth and avoiding beam size modulation resulting from periodic focusing and defocusing fields. In addition to the transverse and longitudinal size matching, the position jitter of the injected beam (~3.1$\mu m$) and the drive laser (~1.3$\mu m$) are kept small compared to their spot sizes to enable a highly collinear overlap of the beam and the wake in space.

Another key to achieving high coupling efficiency is a longitudinally tailored focusing profile to match the beam transverse phase-space. Otherwise, a poor matching may lead to a catastrophic emittance growth, divergence increase and beam loss. In this experiment, the gas jet is specifically designed to produce a plasma structure with a density profile (Fig. 1c) for matching the injected beam to the plasma wake.

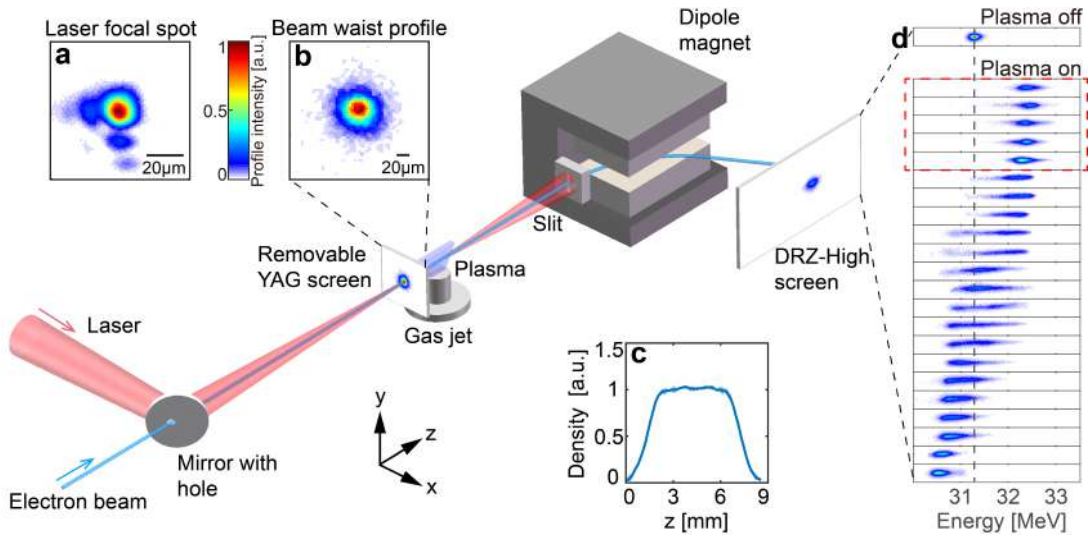

**Figure 1 Experimental layout.** The laser pulse is focused by an $f/12.7$ off-axis parabolic mirror (not shown) and sent collinearly with the electron beam using a mirror with a 3-mm-diameter central hole. The laser focal spot and the electron beam waist profile are shown in inset **a** and **b**, respectively. Inset **c** shows the measured neutral density profile of the gas jet (the blurred region shows the RMS spread of 5 shots). The beam energy spectra are recorded by a spectrometer composed of a permanent dipole magnet of ~1Tesla, a 1mm-wide lead slit and a DRZ-high screen (Mitsubishi Chemical Corporation). The lead slit introduces an uncertainty in the incoming beam position relative to the spectrometer, and thus its width gives an energy measurement uncertainty of ~0.05MeV. Inset **d** shows a group (sorted by decreasing mean energy) of the energy-dispersed beam distributions induced by the ~100-fs timing jitter under the same experimental condition ($z_f = -4.5\ mm$ and $n_p = 6\times10^{17} cm^{-3}$).

In the experiment, by tuning the electron beam arrival time relative to the laser pulse, we can place the beam just in the first few wake wavelengths behind the driver and control the beam injection phase. Since the beam arrival time has a jitter of ~100fs (RMS)[19], energy jitter will be induced in the LWFA, as shown in 22 shots of the energy-dispersed beam distributions (Fig. 1d) measured under the same experimental condition ( $z_f = -4.5\ mm$ and $n_p = 6\times10^{17} cm^{-3}$ ), where both the beam acceleration and deceleration by the plasma wake can be observed. The first five shots of the plasma-on case (inside the red dashed rectangle in Fig. 1d) show features of both maximum energy gain and minimum energy spread, which can be viewed as the electron beam being at the proper acceleration phase of the plasma wakefield.

We vary the focal plane position $z_f$ and the plasma density $n_p$, and find that such monoenergetic acceleration resulting from the proper wake phase is consistent and stable for a certain parameter interval. Examples of the energy-dispersed beam distributions are shown in Fig. 2a-d, where Fig. 2a-c are obtained by decreasing $z_f$ from -3.5mm to -5.5mm while setting $n_p = 6\times10^{17} cm^{-3}$ (Case#1-3) and Fig. 2d is obtained by decreasing $n_p$ to $2\times10^{17} cm^{-3}$ while setting $z_f$ the same as Fig. 2c (Case#4). For comparison, Fig 2e shows one typical shot of the beam distribution without plasma interaction.

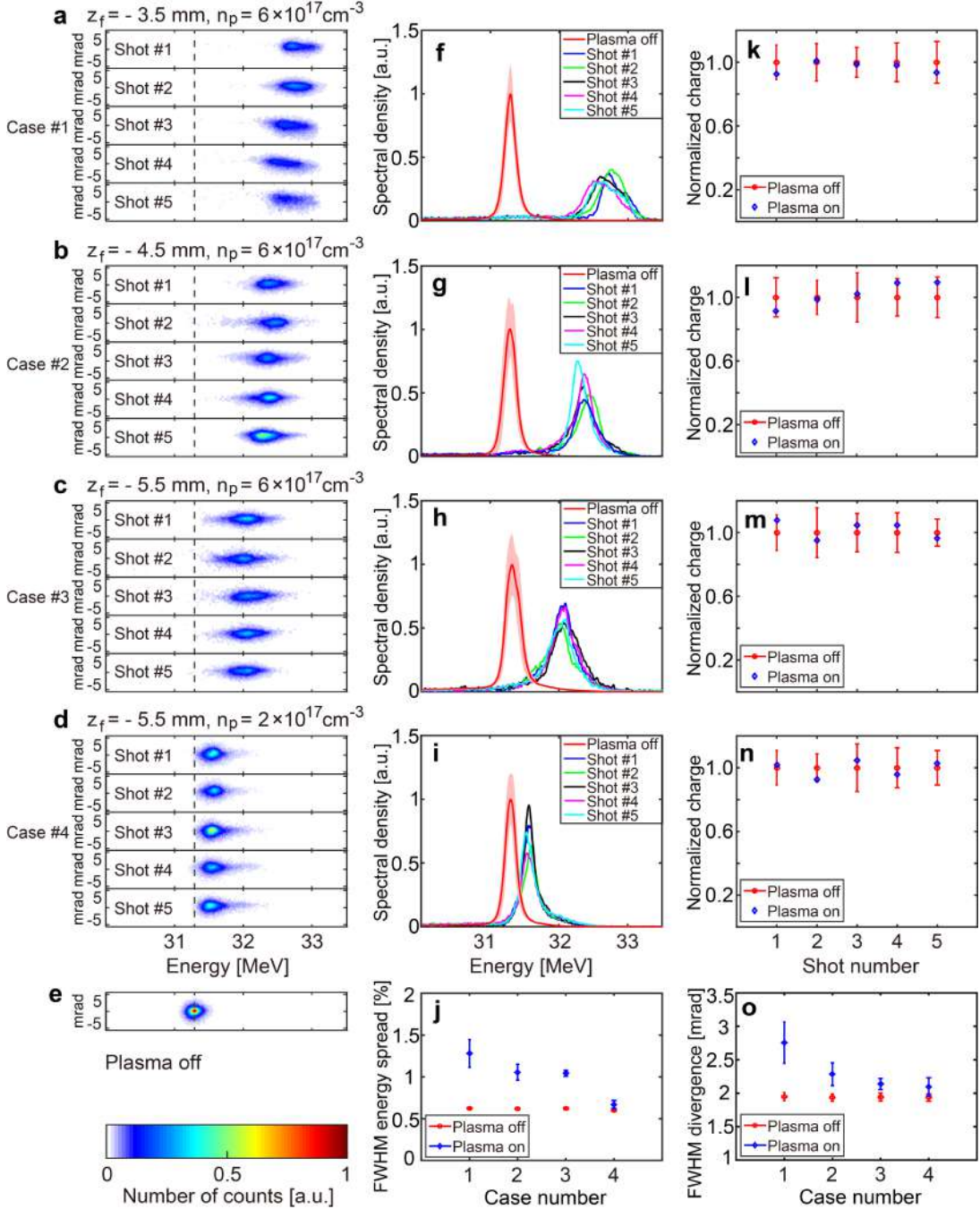

**Figure 2 Experimental results.** The energy-dispersed beam distributions for plasma-on cases with various laser focal plane position $z_f$ and plasma plateau density $n_p$ (**a-d,** 5 shots for each experimental condition), and for the plasma-off case **(e)**. The beam vertical distribution corresponds to the beam transverse divergence since the propagation of the electrons in this direction is not affected by the magnet. **f-i,** the integrated beam energy spectra corresponding to **a-d**. For comparison, 20 consecutive plasma-off shots for each plasma-on shot are simultaneously recorded within ~4 seconds. These total 100 plasma-off shots for each experimental condition (5 plasma-on shots) are shown in **f-i** with red lines, where the blurred regions show the RMS spread of the data. **j**, the beam energy spreads (FWHM) corresponding to **a-d** (5-shot average for the plasma-on case and 100-shot average for the plasma-off case under each experimental condition). **k-n,** the integrated beam charge corresponding to **a-d** (normalized to the average value of 20 plasma-off shots for each plasma-on shot). **o,** the integrated beam divergences (FWHM) corresponding to **a-d**. (5-shot average for the plasma-on case and 100-shot average for the plasma-off case under each experimental condition).

Fig. 2f-i show the integrated energy spectra corresponding to Fig. 2a-d and Fig.2j shows the corresponding average energy spreads. The average peak-to-peak energy gain is ~1.5MeV for the case of $z_f = -3.5mm$ and $n_p = 6\times10^{17} cm^{-3}$. For a plasma length of ~6mm FWHM, this represents an average acceleration gradient of ~250MV/m. The average energy spread is as small as 1.28% FWHM (compared to the initial value of 0.63% FWHM). As $z_f$ decreases, the laser intensity and thus the acceleration gradient decreases since the laser diffracts more (Fig. 2f-h). As $n_p$ decreases, both the acceleration gradient and the phase interval occupied by the beam decrease, therefore the accumulated energy spread also decreases (Fig. 2h-i). Fig. 2k-n show the integrated beam charge corresponding to Fig. 2a-d, respectively. Due

to the near zero launching phase in the photocathode-RF-gun, the bunch charge is sensitive to the jitter of the RF phase and amplitude, leading to a shot-to-shot fluctuation (~±10%) of the bunch charge. For all four cases in Fig. 2k-n, the fraction of captured electrons is nearly 100% (within the ~10% uncertainty of the charge fluctuation). If $z_f$ further increases and $n_p$ remains the same, e.g., $z_f = -1.5mm$ and $n_p = 6 \times 10^{17} cm^{-3}$, the coupling efficiency is reduced to ~40-50%, as shown in Supplementary Fig. 1.

Fig. 2o shows the integrated average beam divergences corresponding to Fig. 2a-d. For the case of $z_f = -3.5mm$ and $n_p = 6 \times 10^{17} cm^{-3}$, the divergence growth at the exit of the LWFA is within ~40% (from initial 1.94mrad to 2.75mrad FWHM), and this value decreases with decreasing $z_f$ or $n_p$.

To reach a deeper understanding of the experiment, full-scale three-dimensional (3D) particle-in-cell (PIC) simulations corresponding to Fig. 2a-d are performed using the code OSIRIS[20,21] (Methods). In Fig. 3a-b, the simulated evolution of the laser spot size $w$ and normalized vector potential $a_0$ ($a_0 = 0.68 \times 10^{-9}\sqrt{I_0[Wcm^{-2}]}$) are shown as a function of the propagation distance $z$, respectively. Due to the laser diffraction in vacuum, $w$ at the entrance of the plasma is several times the electron beam transverse size and it continues to increase during further propagation since the laser power ($P \approx 9TW$) is less than the critical power for relativistic self-focusing ($P_c[TW] = 3 \times 10^{19}/n_p[cm^{-3}]$), resulting in a relatively-low $a_0$ ($a_0 < 1$). Such large $w$ and relatively-low $a_0$ lead to a linear plasma wake[22] with a transverse size much larger than the beam size, which is beneficial for obtaining a high coupling efficiency.

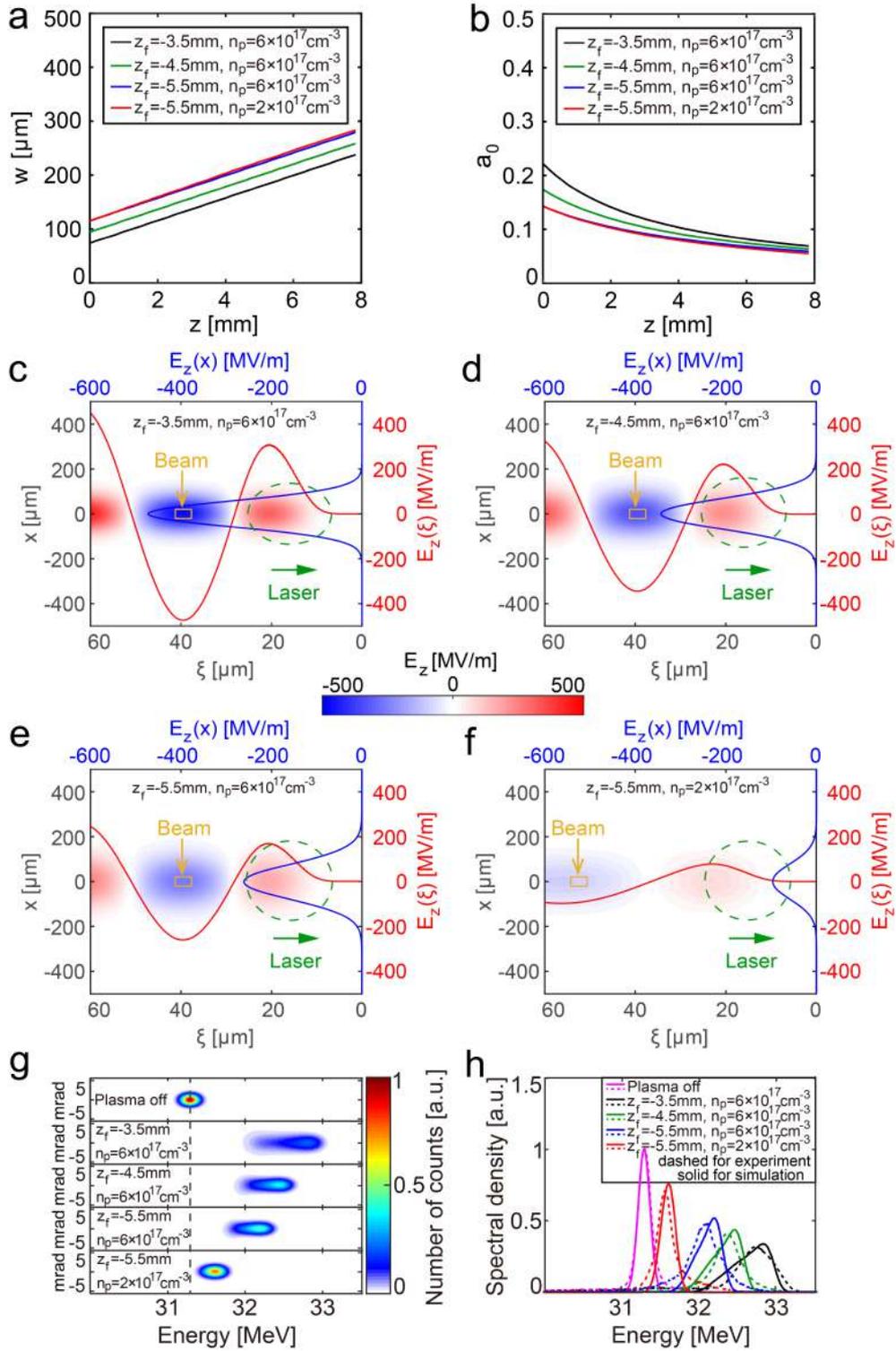

**Figure 3 Simulation results of laser evolution, wakefield structure and output beam energy**

**distribution. a** and **b** show the simulated evolutions of the laser spot size $w$ and the laser vector potential $a_0$ as a function of the propagation distance $z$ in the plasma, respectively. **c-f,** the simulated laser-excited longitudinal wakefields $E_z$ corresponding to **Fig. 2a-d** just at the start of the plasma plateau ($z = 2.4mm$), where $\xi = ct - z$ represents the position relative to the laser pulse. The green and yellow lines show the contours of the laser ($e^{-2}$ of its peak intensity) and the electron beam (RMS bunch length in $\xi$ and RMS bunch size in $x$), respectively. Lineouts of the on-axis $E_z$ and the transverse variation in $E_z$ ($\xi = 39\mu$m for **c-e** and $\xi = 56\mu$m for **f**) are respectively shown with red and blue solid lines. **g,** the simulated angle-resolved energy-dispersed beam distributions. **h,** the simulated beam energy spectra integrated from **g** are shown with solid lines. For comparison, the corresponding experimental results obtained from **Fig. 2f-i** (5-shot average for the plasma-on case under each experimental condition and total 400-shot average for the plasma-off case) are shown with dashed lines.

Fig. 3c-f show the longitudinal wakefields $E_z$ corresponding to Fig. 2a-d right at the start of the plasma plateau ($z = 2.4mm$), where the accelerating gradient is found to be maximum during the whole acceleration process. For efficient acceleration, the beam center is placed near the crest of the acceleration phase, as also shown in Fig. 3c-f. This near on-crest acceleration also leads to the low accumulated energy spread despite no significant beam loading. Simulations confirm that all the electrons can be fully trapped and subsequently accelerated. The resulting energy-dispersed beam distributions and the corresponding energy spectra are shown in Fig. 3g and h, respectively, both in good agreement with the experimental results. Moreover, by varying the delay between the beam and the laser in the simulations, we can obtain the dispersed beam distributions as a function of the delay and a typical example for the case of $z_f = -4.5mm$ and $n_p = 6 \times 10^{17} cm^{-3}$ is shown in Supplementary Fig. 2, very similar to the measured results (Fig. 1d).

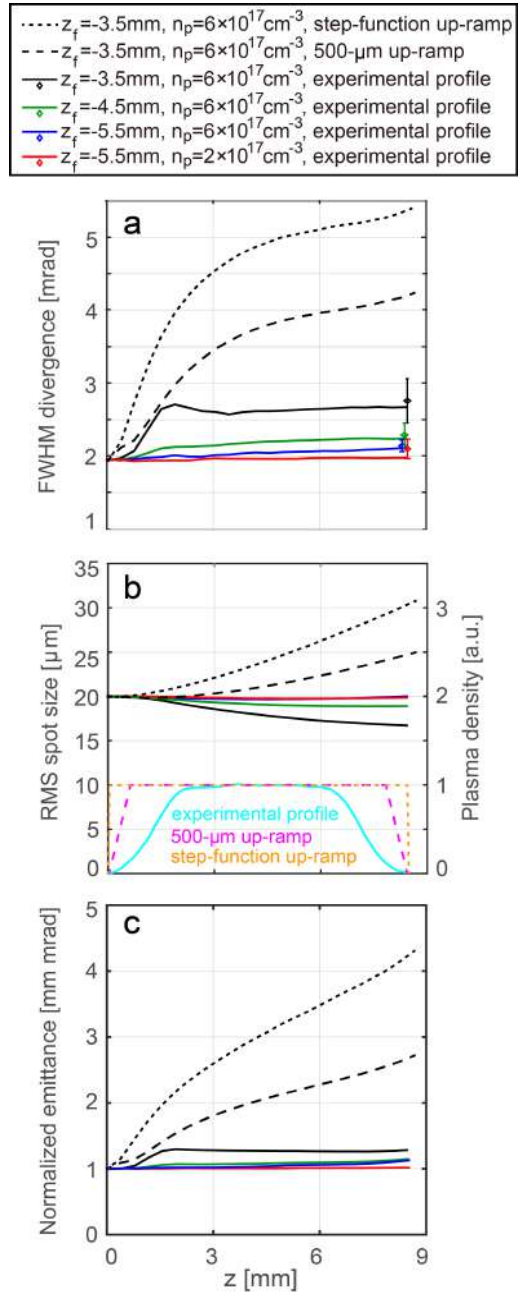

**Figure 4 Simulation results of transverse beam dynamics.** Simulated evolutions of beam divergence (**a**), spot size (**b**) and normalized emittance (**c**) with three different plasma structures (solid lines for the experimental profile, dashed lines for a profile with 500-$\mu m$ up-ramp and dotted lines for a profile with step-function up-ramp) for different $z_f$ and $n_p$ cases. The corresponding measured beam divergences at the exit of the plasma from **Fig.2 o** are also shown in **a** with diamonds.

To get a deeper insight of the beam transverse phase-space dynamics and the possible beam quality degradation during the injection-acceleration process, we plot in Fig. 4 a-c the simulated evolutions of beam divergence, spot size and normalized emittance for different $z_f$ and $n_p$ cases. At the exit of the plasma, the simulated divergences agree well with those measured in the experiment (diamonds in Fig. 4a). The small divergence increases and small variations of the beam spot sizes (Fig. 4b), indicate that the growth of the beam emittance should be limited, and this is confirmed by the simulated emittance evolution in Fig. 4c, where for most cases the emittance growth is only a few percent, with the worst case being ~27% for $z_f = -3.5\ mm$ and $n_p = 6\times10^{17} cm^{-3}$. This level of emittance preservation is indeed highly non-trivial and needs careful tailoring of the plasma profiles. This is because the longitudinal-position-dependent transverse focusing fields in the wake can lead to large phase differences in the betatron oscillation of different beam slices, therefore significant projected emittance growth can be induced if the beam is not properly matched with a carefully chosen plasma profile[14,15] (Supplementary Information). To clearly show the importance of a properly chosen plasma profile, in Fig. 4a-c the simulated evolutions of beam divergence, spot size and normalized emittance for another two different profiles (with a 500-$\mu m$ up-ramp and a step-function up-ramp) are plotted for the case of $z_f = -3.5\ mm$ and $n_p = 6\times10^{17} cm^{-3}$, where a significant growth of the projected emittance (a factor of 2.7 and 4.5 for the 500-$\mu m$ up-ramp and step-function up-ramp cases, respectively) can be seen. In contrast, for the experimental condition where a 2.4mm up-ramp is adopted, the projected emittance growth is fairly small, where the phase differences induced in the plateau have been partially compensated by the phase differences induced in the up-ramp.

In our experiment, a beam charge ~20fC is adopted for achieving a short bunch length ~13fs RMS by velocity compression, where space charge induced longitudinal expansion of the bunch can be mitigated. If the bunch charge is increased by 50 times to 1pC while all other parameters are exactly identical to those in the experiment, simulations confirm that the whole physical process is almost the same. The mono-energetic acceleration with 100% coupling efficiency and beam quality preservation can also be achieved, as clearly shown in Supplementary Fig. 3.

In the current experiment, the laser power is relatively low (~9TW) and the plasma length is much shorter than the dephasing length, especially for $n_p = 2\times 10^{17} cm^{-3}$ case. For this density, a channel guided LWFA has a dephasing length of ~20cm. Using a 40-fs 200-TW laser ($a_0 = 2.2$) interacting with a 20-cm long plasma channel, 3D PIC simulations show that the energy of an injected beam (a 20-pC 25-MeV electron beam with a normalized emittance of 1mm mrad, a focused Gaussian spot of 3.6$\mu m$ RMS and a flat-top current profile of 10fs full duration) can be boosted to ~4.GeV with 100% coupling efficiency and negligible normalized emittance growth (Supplementary Fig. 4). The accelerated beam has a slice energy spread of ~0.06%

RMS with a near linear chirp of ~2.3% RMS, which can be further reduced to ~0.1% level using a low-density plasma dechirper[23-27].

In summary, we have demonstrated experimentally for the first time high efficiency coupling (~100%) and subsequent monoenergetic acceleration between a conventional RF LINAC and a LWFA. This is a crucial milestone for realizing future compact colliders based on plasma accelerators.

## Methods

**Electron beam generation and transport.** A ~20-fC, ~13-fs (RMS), 31.3-MeV electron bunch has been produced by a high-brightness S-band LINAC[16,17] at Tsinghua University. The schematic layout of the beamline is shown in Supplementary Fig. 5a. The bunch charge is set by tuning the energy of the 300-fs (FWHM) photocathode drive laser. The short bunch length is achieved through velocity compression within the photocathode-RF-gun by launching the beam at a near zero phase[18]. High fidelity particle dynamic simulations with the code ASTRA[28] are performed to estimate the bunch length and the beam current profile (near flat-top) (see Supplementary Fig. 5b). After further acceleration in the accelerating tube, the electron beam is transported to the vacuum interaction chamber (experimental area with setup shown in Fig. 1). Two triples are used to focus the beam to the entrance of the plasma with a RMS transverse waist size of $20.3 \pm 0.9 \mu m$, detected by a removable YAG screen in Fig. 1. A 180-nm-thick diamond film (not shown in Supplementary Fig. 5) is inserted between the LINAC and the interaction chamber for differential vacuum pressure. The vertical (y-axis) normalized emittance of the beam after the diamond film is directly measured to be ~1mm mrad by using a two-screen method[29] (YAG screen and DRZ-high screen in Fig. 1).

**Characterization of the plasma.** The plasma structure used in the experiment is produced by a supersonic slit-opening gas jet (6mm×2mm), and its density profile is characterized by combining offline and online measurements with shearing interferometry[30] using a wavefront sensor (SID-4, PHASICS) (see Supplementary Information for details).

**PIC simulations.** Computer simulations are carried out using the 3D fully relativistic PIC code OSIRIS in the moving-window configuration (the simulation box travels at the speed of light in the laser propagation direction).

The experiment-related simulations (presented in Fig. 3, Fig. 4, Supplementary Fig. 2 and Supplementary Fig. 3) are carried out in the laboratory frame. A 3D cylindrical geometry with Fourier azimuthal decomposition[31,32] (the first two Fourier modes) is utilized in these simulations. To make comparisons between the simulations and the experimental results, parameters of the laser, the plasma and the electron beam used in the simulations are chosen as close to the experimental conditions as possible. A 40 fs (FWHM with a $Sin^2$ temporal profile), linearly polarized laser pulse is initialized with $a_0 = 1.35$ and a Gaussian focal spot ($12.2 \mu m$). A transversely uniform plasma is initialized with a measured longitudinal profile as shown in Fig. 1d. A 31.3-MeV electron beam with 0.6% (FWHM) energy spread and 1 mm mrad normalized emittance is initialized with a Gaussian focal spot of $20\mu m$ RMS and a flat-top current profile of 52fs full duration (13fs RMS duration). The simulation window has a dimension of 571.5 μm×63.5 μm with 1500×3000 cells in the $r$ and $z$ directions, respectively. This corresponds to cell sizes of $\Delta r = 3k_0^{-1}$ and $\Delta z = 0.167k_0^{-1}$ (where $k_0 = 2\pi\lambda_0^{-1}$ is the laser wave vector and $\lambda_0 = 800$ nm). 2 macro-particles per cell in the $r-z$ direction and 16 particles in the azimuthal direction are used for both the plasma and the electron beam.

The simulation of the matching section in Supplementary Fig. 4 is performed within the laboratory frame in 3D Cartesian coordinates. The simulation window has a dimension of 228.6 μm×228.6 μm×76.2 μm with 900×900×4800 cells in the $x$, $y$ and $z$ directions, respectively. To save the computational resources, the simulation of the acceleration section in Supplementary Fig. 4 is carried out in a Lorentz-boosted frame[33,34] with the relativistic factor $\gamma_{boost} = 5$ in 3D Cartesian coordinates. An electron beam with almost the same parameters as those at the exit of the matching section is initialized. The simulation window (in the boosted frame) has a dimension of 228.6 μm×228.6 μm×762 μm with 900×900×4800 cells in the $x$, $y$ and $z$ directions, respectively. For simulations of both the matching and the acceleration sections, 1 macro-particle per cell and 8 macro-particles per cell are used for the plasma and the electron beam, respectively.

# References

1. Tajima, T. & Dawson, J. M. Laser electron accelerator. *Phys. Rev. Lett.* **43,** 267–270 (1979).

2. Chen, P. *et al.* Acceleration of electrons by the interaction of a bunched electron beam with a plasma. *Phys. Rev. Lett.* **54,** 693–696 (1985).

3. Faure, J. *et al.* A laser–plasma accelerator producing monoenergetic electron beams. *Nature* **431,** 541–544 (2004).

4. Geddes, C. G. R. *et al.* High-quality electron beams from a laser wakefield accelerator using plasma-channel guiding. *Nature* **431,** 538–541 (2004).

5. Mangles, S. P. D. *et al.* Monoenergetic beams of relativistic electrons from intense laser–plasma interactions. *Nature* **431,** 535–538 (2004).

6. Blumenfeld, I. *et al.* Energy doubling of 42 GeV electrons in a metre-scale plasma wakefield accelerator. *Nature* **445,** 741–744 (2007).

7. Corde, S. *et al.* Muiti-gigaelectronvolt acceleration of positrons in a self-loaded plasma wakefield. *Nature* **524,** 442–445 (2015).

8. Steinke, S. *et al.* Multistage coupling of independent laser-plasma accelerators. *Nature* **530,** 190–193 (2016).

9. Adli, E. *et al.* Acceleration of electrons in the plasma wakefield of a proton bunch. *Nature* **561,** 363–367 (2018).

10. Litos, M. *et al.* High-efficiency acceleration of an electron beam in a plasma wakefield accelerator. *Nature* **515,** 92–95 (2014).

11. The Particle Physics Project PrioritizationPanel (P5). Subpanel of the High

Energy Physics Advisory Panel (HEPAP). *Building for Discovery, Strategic Plan for U.S. Particle Physics in the Global Context.* http://www.usparticlephysics.org/p5/ (2014).

12. Schroeder, C. B., Esarey, E., Geddes, C. G. R., Benedetti, C. &Leemans, W. P. Physics considerations for laser-plasma linear colliders. *Phys. Rev. Spec. Top. - Accel. Beams* **13**, 101301 (2010).

13. Adli, E. *et al.* A beam driven plasma-wakefield linear collider: fromHiggs factory to multi-TeV. *In Proc. Community Summer Study 2013: Snowmass on the Mississippi.* Preprint at http://arxiv.org/abs/1308.1145 (2013).

14. Xu, X. L. *et al*. Physics of phase space matching for staging plasma and traditional accelerator components using longitudinal tailored plasma profiles. *Phys. Rev. Lett.* **116,** 124801 (2016).

15. Dornmair, I. *et al.* Emittance conversation by tailored focusing profiles in a plasma accelerator. *Phys. Rev. Spec. Top. - Accel. Beams* **18,** 041302 (2015).

16. Du, Y. C. *et al*. Generation of first hard X-ray pulse at Tsinghua Thomson Scattering X-ray source. *Rev. Sci. Instrum.* **84,** 053301 (2013).

17. Zheng, L. M. *et al*. Development of S-band photocathode RF guns at Tsinghua University. *Nucl. Instr. Meth. Phys. Res. Sect. A* **834,** 98-107 (2016).

18. Zhang, Z. *et al.* High time resolution beam-based measurement of the rf-to-laser jitter in a photocathode rf gun. *Phys. Rev. Spec. Top. - Accel. Beams* **17,** 032802 (2014).

19. Lin, Z. Y. *et al.* Development of sub-100 femtosecond timing and

synchronization system. *Rev. Sci. Instrum.* **89,** 014701 (2018).

20. Fonseca, R. A. *et al.* OSIRIS: A Three-Dimensional, Fully Relativistic Particle in Cell Code for Modeling Plasma Based Accelerators. in *Computational Science — ICCS 2002* (ed. P. Sloot, A. Hoekstra, Dongarra, C. T. and J. D.) 342–351 (Springer Berlin / Heidelberg, 2002).

21. Fonseca, R. A. *et al.* One-to-one direct modeling of experiments and astrophysical scenarios: pushing the envelope on kinetic plasma simulations. *Plasma Phys. Control. Fusion* **50,** 124034 (2008).

22. Esarey, E., Schroeder, C. B. & Leemans, W. P. Physics of laser-driven plasma-based electron accelerators. *Rev. Mod. Phys.* **81,** 1229-1285 (2009).

23. Wu, Y. P. *et al*. A preliminary experimental study of energy chirp reduction by a plasma dechirper. in *Proceedings of IPAC17, Copenhagen, 2017,* pp. 1258-1260.

24. Wu, Y. P. *et al.* Phase space dynamics of a plasma wakefield dechirper for energy spread reduction. *Phys. Rev. Lett.* **122**, 204804 (2019).

25. Wu, Y. P. *et al.* Near-ideal dechirper for plasma-based electron and positron acceleration using a hollow channel plasma. *Phys. Rev. Applied.* **12**, 064011 (2019).

26. D'Arcy, R. *et al.* Tunable plasma-based energy dechirper. *Phys. Rev. Lett.* **122**, 034801 (2019).

27. Shpakov, V. *et al.* Longitudinal phase-space manipulation with beam-driven plasma wakefields. *Phys. Rev. Lett.* **122**, 114801 (2019).

28. Floettman, K. *http://www.desy.de/~mpyflo/* page ASTRA code.

29. Zhang, L. Y. *et al*. Calculation of two-screen emittance measurement. *Nucl. Instr. Meth. Phys. Res. Sect. A* **407,** 356-358 (1998).

30. Chanteloup, J.-C. Multiple-wave lateral shearing interferometry for wave-front sensing. *Appl. Opt.* **44,** 1559–1571 (2005).

31. Lifschitz, A. F. *et al.* Particle-in-Cell modelling of laser-plasma interaction using Fourier decompostion. *J. Comput. Phys.* **228,** 1803–1814 (2009).

32. Davidson, A. *et al.* Implementation of a hybrid particle code with a PIC description in $r - z$ and a gridless description in $\phi$ into OSIRIS. *J. Comput. Phys.* **281,** 1063–1077 (2015).

33. Martins, S. F. *et al.* Exploring laser-wakefield-accelerator regimes for near-term lasers using particle-in-cell simulation in Lorentz-boosted frames. *Nature Physics* **6,** 311–316 (2010).

34. Martins, S. F. *et al.* Numerical simulations of a laser wakefield accelerators in optimal Lorentz frames. *Comput. Phys. Comm.* **181,** 869–875 (2010).

**Data availability.**

The data that support the findings of this study are available from the corresponding authors on reasonable request.

**Acknowledgements**

This work is supported by the National Natural Science Foundation of China (NSFC) Grants (No. 11535006, No. 11991071, No. 11775125, and No. 11875175), CAS Center for Excellence in Particle Physics, and the U.S. Department of Energy Grants (No. DE-SC0010064, No. DE-SC0008491, and No. DE-SC0008316) at UCLA.

**Author contributions**

W.L. conceived and supervised the project. J.F.H. led the development of laser system, plasma structure and diagnostics, Y.C.D. led the optimization of electron beam, Y.P.W, Z.Z., S.L., B.P., J.F.H., Y.C.D. and W.L. performed the experiments. Y.P.W. carried out the corresponding simulations. W.L., C.J., Y.P.W., J.F.H wrote the paper. All authors contributed extensively to the work presented in this paper.

**Competing interests**

The authors declare no competing interests.

**Additional Information**

Correspondence and requests for materials should be addressed to W.L. (weilu@tsinghua.edu.cn) or J.F.H. (jfhua@tsinghua.edu.cn) or Y.C.D. (dych@tsinghua.edu.cn).

# Supplementary Information

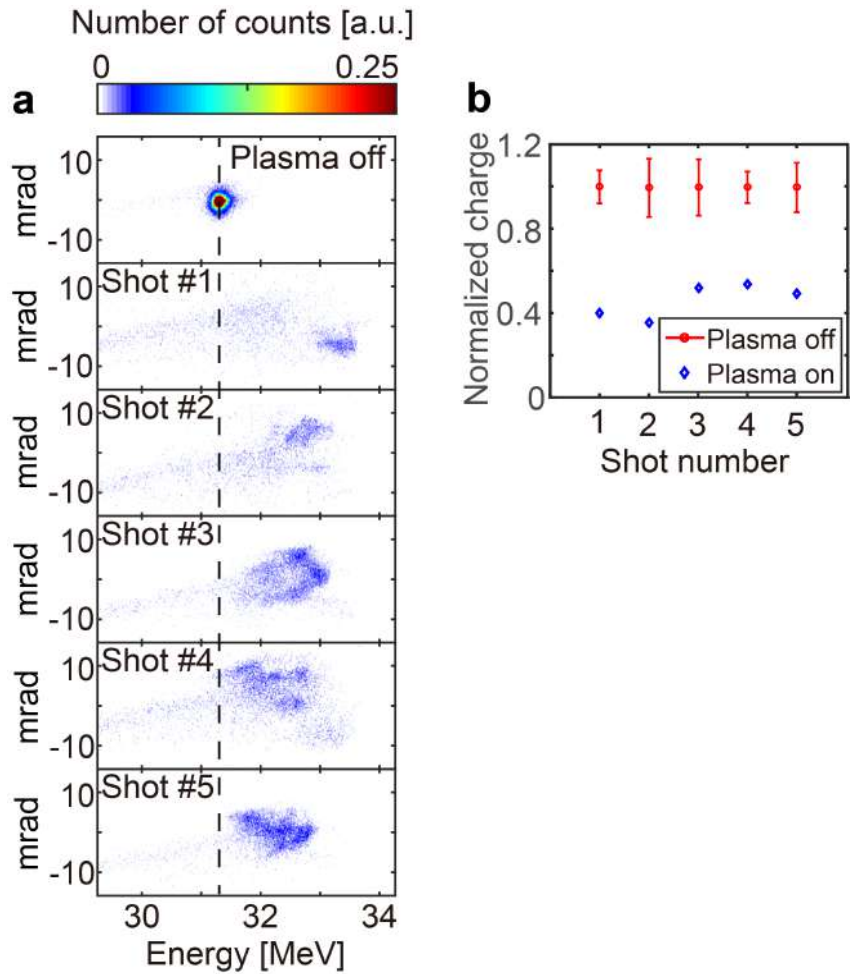

**Supplementary Figure 1 Experimental cases for non-matched injection with low coupling efficiency. a,** the measured energey-dispersed beam distributions, with $z_f$ = -1.5 mm and $n_p = 6\times10^{17}$cm$^{-3}$ . **b,** the integrated normalized beam charge corresponding to **a** (20-plasma-off-shot average for each plasma-on shot).

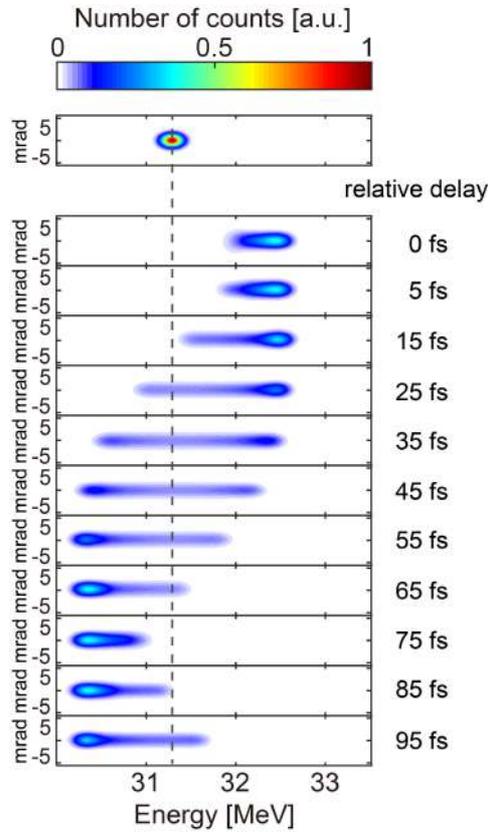

**Supplementary Figure 2 Simulated energy-dispersed beam distributions as a function of the relative delay between the electron beam and the laser, with** $z_f = -4.5$ mm **and** $n_p = 6\times10^{17}$ cm$^{-3}$**.** Here the relative delay of 0 fs corresponds to the absolute delay as shown in **Fig. 3d** (the beam center is placed near the crest of the acceleration phase in the first wake bucket).

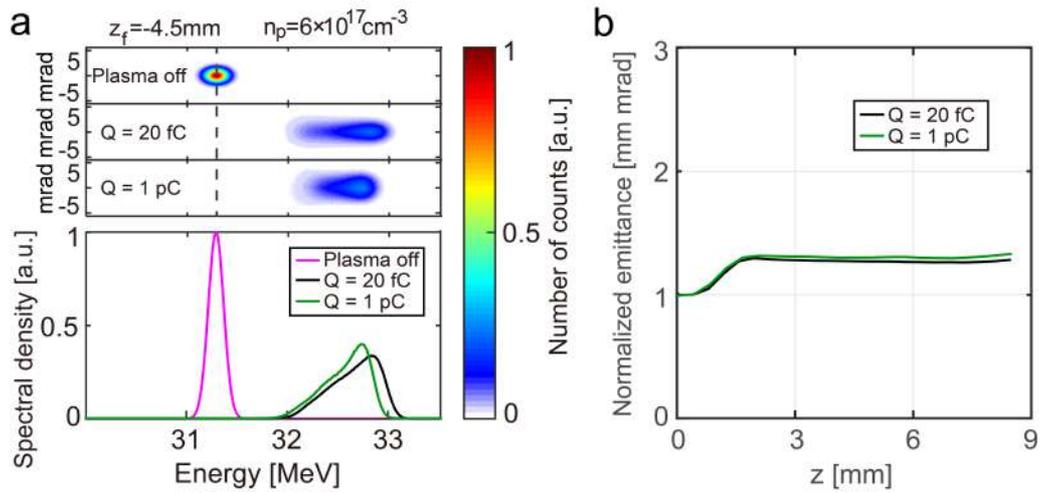

**Supplementary Figure 3 Simulation results obtained by increasing the beam charge to 1 pC while keeping other parameters invariant, with $z_f = -3.5$ mm and $n_p = 6\times10^{17}$ cm$^{-3}$.** a, the simulated angle-resolved energy-dispersed beam distributions and corresponding energy spectra. b, the simulated evolutions of beam normalized emittance.

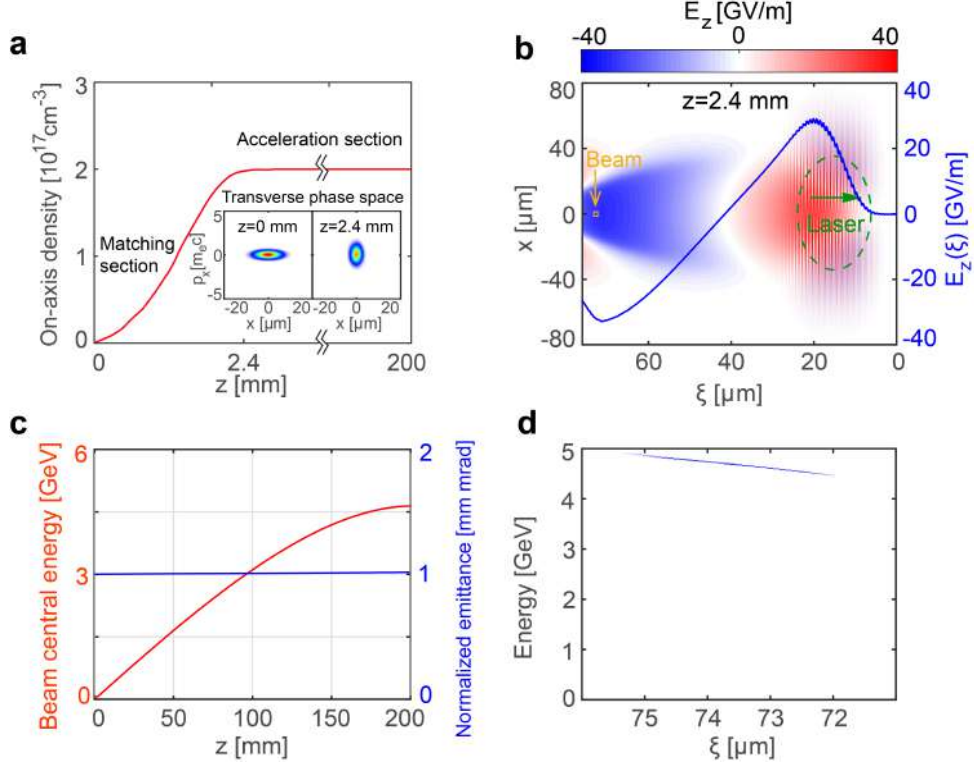

**Supplementary Figure 4 Simulation results for high energy gain when using a longer plasma and a more powerful laser.** The plasma has a 2.4mm-long up-ramp identical to the experimental condition as the matching section, and a ~20cm-long plateau as the acceleration section (see **a**). A 200-TW laser is focused to a spot size $w_0 = 35\mu m$ with $a_0$=2.2 at the beginning of the plateau ($z = 2.4$mm). The plasma transverse profile is set to a parabolic channel $n_{p,0} \times (1 + 0.4 \times \frac{x^2+y^2}{w_0^2})$ for laser guiding, where $n_{p,0}$ is the on-axis density with the plateau value of $2\times 10^{17} \text{cm}^{-3}$. A 20-pC, 25-MeV, 10-fs-full-duration (flat-top current profile) electron beam with 0.6% FWHM (no chirp) energy spread and 1mm mrad normalized emittance is focused to $z = 0$ with a transverse waist size of $3.6\ \mu m$ RMS (left inset in **a**). The matching section can transport the beam from this waist to another waist with spot size of $\sim 1.5 \mu m$ RMS and energy of ~50 MeV at z=2.4 mm (right inset in **a**), which is nearly matched to the plasma focusing fields in the acceleration section. **b,** The simulated $E_z$ at $z = 2.4$mm. The blue line shows the lineout of the on-axis value. The green and yellow lines show the contours of the laser ($e^{-2}$ of its peak intensity) and the electron

beam (RMS bunch length in $\xi$ and RMS spot size in x), respectively. **c,** The evolutions of the beam central energy (red line) and normalized emittance (blue line). **d,** The final beam longitudinal phase space.

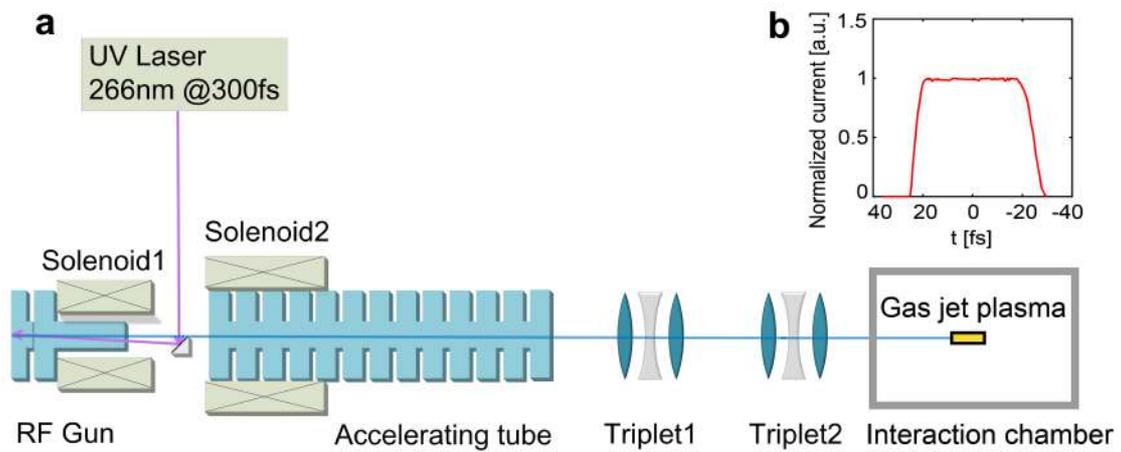

**Supplementary Figure 5 Ultrashort electron beam generation and transport.** a, schematic layout of the LINAC beamline. b, simulated beam current profile (charge 20 fC) according to the experimental settings, where the beam head locates at the right.

**Transverse beam dynamics.**

In a linear plasma wake, the transverse force felt by an electron in the beam can be expressed as $F_\perp = -e(E_r - cB_\theta) \propto -e \frac{a_0^2 k_p r}{w^2} \exp\left(-\frac{2r^2}{w^2}\right) \sin \Psi$, where $\Psi = k_p(\xi - \xi_l)$ is the injection phase with the position of the drive laser $\xi_l$. Since in our experiment the beam size is much smaller than the laser transverse size in the plasma, i.e., $r \ll w$, the exponential term can be omitted and thus $F_\perp$ is linear in $r$ in the beam region. This force linearity can preserve the beam slice emittance. In the focusing region, the individual particles perform transverse betatron oscillations with a betatron frequency of $\omega_\beta = \sqrt{\frac{|F_\perp|}{r \gamma_b m_e}}$, where $\gamma_b$ is the beam Lorentz factor and $m_e$ is the electron rest mass.

Due to the relatively long bunch length or finite beam energy spread, electrons at different $\xi$ or with different $\gamma_b$ oscillate at different $\omega_\beta$. For our experimental parameters ($\delta_\xi \gg \delta_{\gamma_b}$), the $\xi$-dependence dominates the $\omega_\beta$-variance. Thus large difference can be induced in the electron betatron phase advance of various longitudinal bunch slices, leading to a growth in beam projected emittance.

To prevent the emittance growth, the transverse properties (beta function or the transverse waist size) of the beam should be matched to the intrinsic electron betatron motion in the plasma wake. Nevertheless, since the betatron phase advance is $\xi$-dependent in a linear wake, the matching condition cannot be satisfied for the whole beam except for a certain slice. In our experiment, in order to optimize the overall matching effect for the beam, this certain slice is approximately chosen to be the beam center slice. In addition, to suppress the betatron decoherence of the other slices, a specifically designed plasma profile is adopted in the experiment.

Taking the case of $z_f = -3.5\ mm$ and $n_p = 6\times10^{17} cm^{-3}$ as an example, a quasi-matching condition has been achieved in our experiment with a proper beam transverse waist size ($\sigma_r = 20\mu m$). If $\sigma_r$ is far from the quasi-matching condition (i.e. $\sigma_r = 10\mu m$, $30\mu m$ and $40\mu m$, with other parameters unchanged), significant projected emittance growth can be induced, as shown in Supplementary Fig. 6.

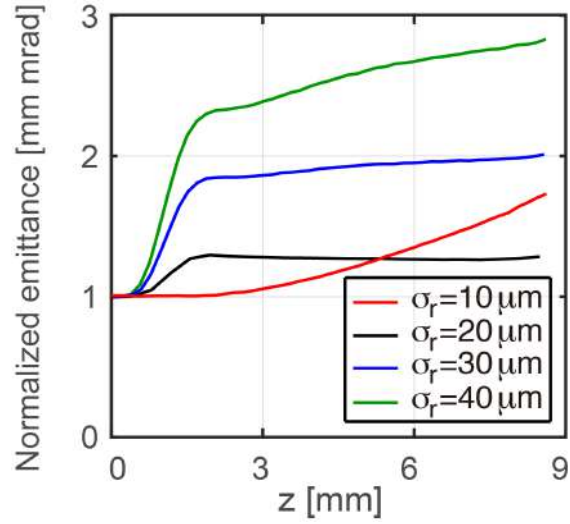

**Supplementary Figure 6** Simulated evolutions of beam normalized emittance for four different beam transverse waist sizes ($\sigma_r = 10\mu m$, $20\mu m$, $30\mu m$ and $40\mu m$), with $z_f = -3.5\ mm$ and $n_p = 6\times10^{17} cm^{-3}$.

The properly designed plasma profile plays an important role in reduction of the electron betatron decoherence. When the beam enters the low-density region of the plasma upramp where the plasma wavelength is relatively large, the beam stays in the focusing phase with a positive slope $\partial(E_r - cB_\theta)/\partial\xi > 0$, leading to a faster phase ellipse rotation speed of the bunch tail than that of the bunch head in the transverse phase space. As the beam propagates to the following plasma plateau, $\partial(E_r -$ $cB_\theta)/\partial \xi$ changes the sign to < 0 and thus the phase ellipse of the bunch tail rotates slower than that of the bunch head in the transverse phase space, which can partially compensate previously accumulated betatron phase advance differences of each beam slice due to $\partial(E_r - cB_\theta)/\partial \xi > 0$. This effect can suppress further projected beam emittance growth (see Fig. 4c, where a comparison of the beam emittance evolution with the experimental profile and another two profiles is shown).

**Characterization of the plasma structure**

The plasma structure used in the experiment is produced by a supersonic slit gas jet (6mm×2mm, see Supplementary Fig. 7a), and its density distribution is fully characterized by multiplying the offline-measured normalized density profile and the directly measured absolute electron density of the plateau. Both measurements are based on shearing interferometry with a wavefront sensor (SID-4, PHASICS).

In the offline measurement, argon gas is used to measure the gas profile for its relatively large refractive index. Fluid simulations and previous experiments confirm that the normalized density profiles for different gases (Ar, He) are similar at a given backing pressure[1]. A typical phase map of the gas flow is shown in Supplementary Fig. 7b. For a gas jet with rectangular slit cross section, the normalized argon density profile along the beam path is approximately equal to the corresponding normalized phase map profile. The experimental data confirm that the maximum gas density is linearly proportional to the backing pressure, and its normalized profile is nearly invariant (Supplementary Fig. 7c).

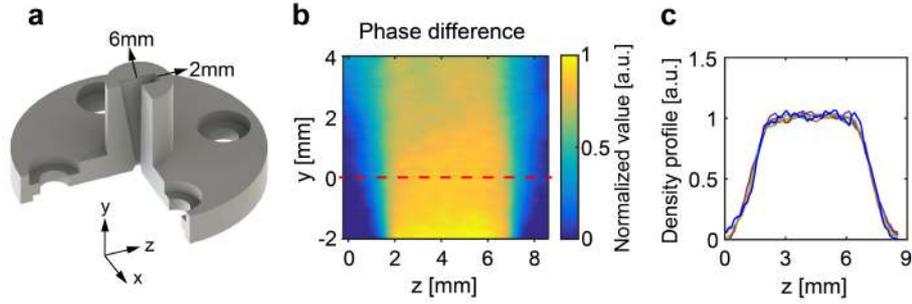

**Supplementary Figure 7  Gas jet and offline measurement results. a,** the three-dimensional diagram of the gas jet. The drive laser and the electron beam propagate along the z axis. **b,** a typical phase map of the argon gas above the gas jet. The top edge of the jet is located at $y = -2mm$ and the beam path is located at $y = 0mm$ (the red dashed line). **c,** the longitudinal normalized argon density profile along the beam path under different gas pressures (lines with different colors).

In the experiment, helium gas is used, and a direct measurement of the plasma electron density can be obtained through interferometry using a pulse splitting off from the ionization laser (power chosen to fully ionize the helium gas on axis). The delay between the two pulses is ∼ 50 ps, short enough to avoid any plasma density evolution. The phase maps near the middle of the plasma plateau (limited by the view of the wavefront sensor) under different gas pressures $P_g$ (Supplementary Fig. 8a for $P_g = 1.5$ MPa) are obtained for the plasma electron density retrieval using Abel inversion (Supplementary Fig. 8b for $P_g = 1.5$ MPa). The mean plasma density on axis varies almost linearly with $P_g$ (Supplementary Fig. 8c).

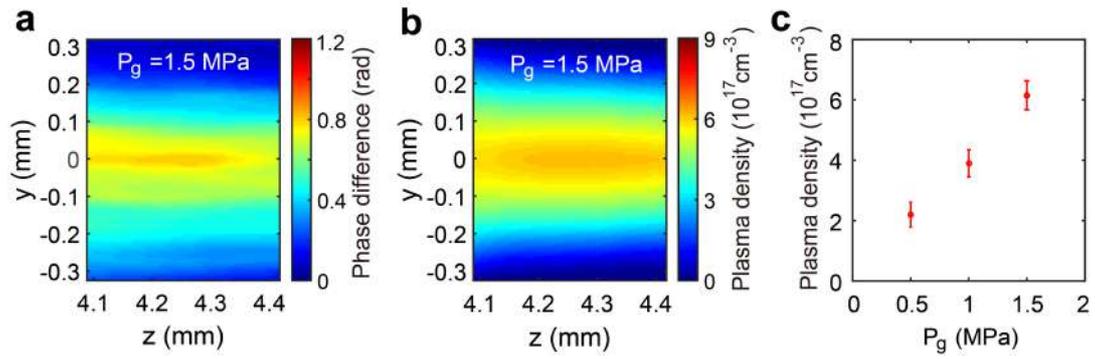

**Supplementary Figure 8    Online measurement results. a,** phase map image of $P_g = 1.5$ MPa. **b,** retrieved plasma density image of $P_g = 1.5$ MPa. **c,** the mean plasma density on-axis versus the gas pressure $P_g$.

# References

1.  Couperus, J.P. *et al.* Tomographic characterisation of gas-jet targets for lasre wakefield acceleration. *Nucl. Instr. Meth. Phys. Res. A* **830,** 504–509 (2016).